\documentclass[10pt,a4paper,pre,twocolumn]{revtex4}

\usepackage{amsmath}
\usepackage{amsfonts}
\usepackage{amssymb}
\usepackage{graphicx}
\begin{document}
\title{Granular packs under vertical tapping: structure evolution, grain motion, and dynamical
heterogeneities}

\author{Massimo Pica Ciamarra}
\email[]{picaciam@na.infn.it}
\affiliation{Dipartimento di Scienze Fisiche, Universit\'a di
Napoli `Federico II' and INFM, Unit\'a di Napoli, 80126 Napoli, Italia.}
\author{Mario Nicodemi}
\affiliation{Department of Physics, University of Warwick, Coventry CV4 7AL, UK}
\author{Antonio Coniglio}
\affiliation{Dipartimento di Scienze Fisiche, Universit\'a di
Napoli `Federico II' and INFM, Unit\'a di Napoli, 80126 Napoli, Italia.}
\homepage{http://smcs.na.infn.it}

\begin{abstract}
The compaction dynamics of a granular media subject to a sequence of vertical taps made of fluid pulses is
investigated via Molecular Dynamics simulations. Our study focuses on three different levels: macroscopic
(volume fraction), mesoscopic (Vorono\"{\i} volumes, force distributions) and microscopic (grain
displacements). We show that the compaction process has many characteristics which are reminiscent of the
slow dynamics of glass forming systems, as previously suggested. For instance the mean volume fraction
slowly increases in time and approaches a stationary value following a stretched exponential law, and the
associated compaction time diverges as the tapping intensity decreases. The study of microscopic quantities
also put in evidence the existence of analogies with the dynamics of glass formers, as the existence of
dynamical heterogeneities and spatially correlated motion of grains; however it also shows that there are
important qualitative differences, as for instance in the role of the cage effect. Correlations between
geometry and dynamics of the system at the grain level are put in evidence by comparing a particle
Vorono\"{\i} volume and its displacement in a single tap.
\end{abstract}
\maketitle

\section{Introduction}
When subject to vertical vibrations granular materials can produce a
variety of distinct phenomena, depending both on the driving
parameters and on the container properties. A great deal of interest
has been recently raised by the process of compaction under
successive vertical taps~
\cite{Nagel95,Nowak98,Rosato00,Philippe02,Swinney05}.
Experiments~\cite{Nagel95, Nowak98, Philippe02} show that the
density of a column of grains submitted to vertical tapping
increases slowly in time with a law well fitted by a stretched
exponential~\cite{Philippe02} or logarithmic
function~\cite{Nagel95}, and that the characteristic compaction time
grows abruptly as the driving intensity decreases. These
observations have suggested an analogy with glass-forming systems,
where the relaxation time diverges as the temperature is decreased;
the analogy is corroborated by the fact that concepts like
frustration and free volume, which are commonly used to explain the
slow dynamics of supercooled liquids and other thermal systems, do
also provide an insight in the physics of powder compaction~
\cite{Mimmo97,Nowak98,Ben-Naim}.

However, contrary to supercooled liquids, granular materials are non-thermal systems as their typical
energy scale, the energy required to rise a grain of its own diameter against gravity, is orders of
magnitude larger than the thermal energy $k_BT$: each granular pack is in a mechanically stable state,
which last as long as there is no external perturbation. Therefore even though the dynamics induced by a
sequence of vertical taps becomes slower and slower as the tapping intensity decreases, important qualitative differences may
exist between slow granular dynamics and glassy dynamics.

Here we investigate analogies and differences between granular
dynamics and glassy dynamics by performing molecular dynamics (MD)
simulations of a granular system subject to a sequence of vertical
taps where, in order to explore a wide range of volume fractions,
the system is tapped via flow pulses as in the experiment of
Schr$\ddot{\rm o}$ter {\it et. al.}~\cite{Swinney05}. We
investigate the evolution of macroscopic quantities (volume
fraction), mesoscopic quantities (Vorono\"{\i} volumes, force
distributions) and microscopic quantities (grain displacements), and
we discuss how various static properties of a granular pack change
during compaction~\cite{Richard03}. We found severals analogies
between the compation of granular media, 
and the slow dynamics of glass forming systems~\cite{Coniglio,Liu,fierro1}, as the divergence of the
relaxation times, dynamical heterogeneities and spatially correlated
motion; but we also show evidences of important qualitative
differences, as in the role of the cage motion, which are put in
evidence by the study of particle trajectories.

The paper is organized as follows. Sec.~\ref{sec-model} presents the numerical model used. Then compation
dynamics, investigated via the study of the time dependence of the volume fraction, and of the diffusion coefficient at stationarity, is discussed in Sec.~\ref{sec-cd}. The evolution of structural properties
of a compacting granular pack, the radial distribution function, the distribution of Vorono\"{\i} volumes,
and the distribution of interparticle forces, is presented in Sec.~\ref{sec-str}. Sec.~\ref{sec-st}
discusses the compaction dynamics at a grain level, showing the existence of dynamical heterogeneities
and of spatially correlated motion of grains, and putting in evidence qualitative differences in the
particle trajectories of compacting granular media and glass formers. Sec.~\ref{sec-st} also presents a
connection between geometrical (Vorono\"{\i} volumes) and dynamical (particle displacements) properties
of the  system. Finally, a conclusion summarizes the main results and perspectives.

\section{Numerical model\label{sec-model}}
We run Molecular Dynamics (MD) simulations of $N=1600$ monodisperse
spherical grains of diameter $d = 1$cm and mass $m=1$g. Grains, under gravity,
are confined in a box with a square basis of length $L = 10$cm,
with periodic boundary conditions in the horizontal directions.
The bottom of the box is made of other immobile, randomly displaced,
grains (to prevent crystallization). Simulations with $4$ time more particles
in a system with a square basis of length $2L$ give the same results,
so we exclude the presence of finite size effect in our system.

Two grains in contact interact via a normal
and a tangential force. The former is given by the
spring-dashpot model, while the latter is implemented
by keeping track of the elastic shear displacement throughout the
lifetime of a contact~\cite{Silbert1}.
The model is the one described in \cite{modello} with
a restitution coefficient, $e= 0.8$.
We use the linear model insted of the more realistic Hertzian model~\cite{hertz}
as this latter is characterized by a coefficient of restitution
which goes to zero as the relative velocity of the impacting particles
decreases~\cite{Silbert1}. This feature makes computationally expensive the simulation
of a granular system reaching a mechanically stable state.

\begin{figure}[t]
\begin{center}
\includegraphics*[scale=0.8]{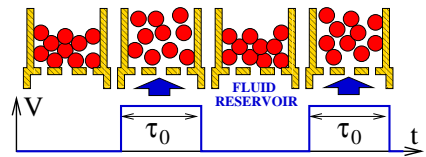}
\end{center}
\caption{\label{f_model} (Color online) We study the compaction
process of a granular media subject to flow pulses. During each
pulse (tap), of duration $\tau_0$, fluid flows with velocity $V$
trough the granular media. Before applying a flow pulse we wait
until the granular media comes to a rest.}
\end{figure}

The system is immersed in a fluid and, starting from a random
configuration, it is subject to a dynamics made of a sequence of
flow pulses where the fluid flows through the grains (see Fig.\ref{f_model}),
as in the experiment of Ref.\cite{Swinney05}.
In a single pulse the flow velocity, directed against gravity, is $V>0$
for a time $\tau_0$; then the fluid comes to rest.
We model the fluid-grain interaction as in Ref.s~\cite{King,Crowe}
via a viscous force proportional to the fluid grain relative velocity:
${\bf F}_{fg}= -A({\bf v} - {\bf V})$
where ${\bf v}$ is the grain and ${\bf V}=(0,0,V)$ is the fluid velocity. The prefactor
$A={\gamma}{(1-\Phi_l)^{-3.65}}$ is dependent on the local
packing fraction, $\Phi_l$,
in a cube of side length $3d$ around the grain, and
the constant is $\gamma = 1$ Ns/cm \cite{Crowe}.

During each pulse, grains are fluidized and then come to rest under
the effect of gravity $g$. The system is considered to be still when
the kinetic energy per grains is below $10^{-5} mgd$.
All measures below are recorded when the pack is at rest.

The dynamics of dry granular media subject to vertical vertical vibrations
is determined by two parameters, the amplitude and the
frequency of vibrations. In the system we are investigating here
there are also two parameters, the tap duration $\tau_0$ and the
fluid velocity $V$.

\section{Dynamics\label{sec-cd}}
\subsection{Volume fraction\label{sec-phi}}
\begin{figure}[t]
\begin{center}
\includegraphics*[scale=0.32]{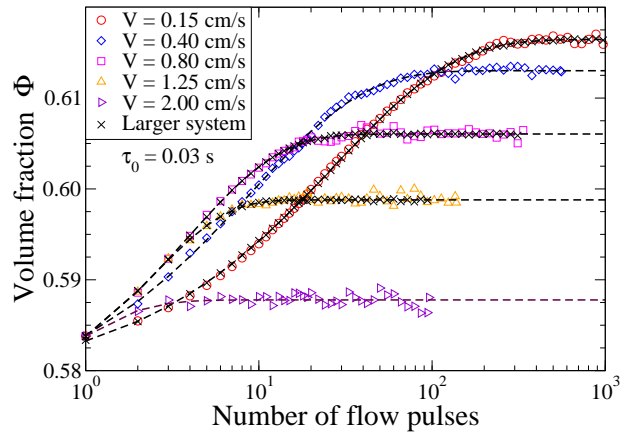}
\end{center}
\caption{\label{f_compaction} (Color online) Temporal evolution of the mean volume fraction
$\Phi$ for $\tau_0 = 0.03$~s and the reported values of fluid velocities $V$. The data obtained via
simulations of a larger systems ($4$ times more particles) evidence the absence of finite size effects.
Dashed lines
are fit to a stretched exponential law.
}
\end{figure}

When subject to a sequence of flow pulses a granular system compactifies (or expands)
until it reaches a stationary state which depends on the driving parameters~\cite{Swinney05}.
Fig.~\ref{f_compaction} shows the evolution of the volume fraction $\Phi$ (measured in the bulk of the
system) with the number of flow pulses for systems subject to a tap dynamics with tap length
$\tau_0 = 0.03$s and various values of the fluid velocity $V$. Each curve is obtained by averaging
over $32$ independent realizations. Similar curves are obtained with different values of $\tau_0$.

The time evolution of the volume fraction is well described by a stretched exponential law,
\begin{equation}
\label{eq-se}
\Phi(t) = \Phi_\infty - (\Phi_\infty -  \Phi_0) \exp\left[-(t/\tau)^c\right],
\end{equation}
with $c \simeq 1$.
This is in agreement with experimental results of Philippe {\it et. al.}
\cite{Philippe01, Philippe02, Coniglio}, which have investigated the relaxation dynamics of dry granular
media subject to vertical taps in a system with a height to width ratio similar to ours. On the contrary
Nowak {\it et. al.}~\cite{Nagel95, Nowak98} have investigated a system with a larger height to width ratio,
finding logarithmic compaction.

\begin{figure}[t!]
\begin{center}
\includegraphics*[scale=0.32]{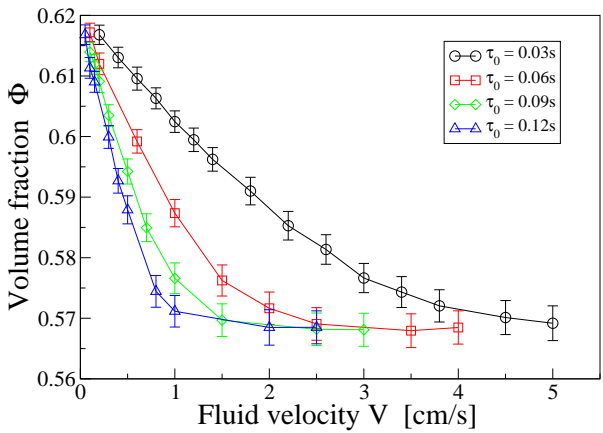}
\end{center}
\caption{\label{f_phi_u} (Color online) Dependence of the volume fraction reached at stationarity by a
system subject to a sequence of flow pulses of length $\tau_0$ on the fluid velocity $V$. Lines are a
guide to the eye.
}
\end{figure}

As in previous experiments of both
vibrated~\cite{Ben-Naim,Nowak97,Nowak98,Philippe02} and
fluidized~\cite{Swinney05} granular systems, when the tapping
intensity decreases the system compactifies more. This is clearly
illustrated in Fig.~\ref{f_phi_u}, where we show the dependence of
the volume fraction reached at stationarity on the fluid velocity
$V$ for various values of the tap length $\tau_0$. As one could have
expected the final stationary state depends both on $V$ and $\tau_0$. 
However it is possible to show numerically~\cite{noi_slow_rev} that
the final state can be characterized by one thermodynamical
parameter, supporting the idea of a statistical mechanics
description of granular media at rest~\cite{fierro,barrat,makse,Coniglio} 
originally proposed by S.F. Edwards~\cite{Edwards}.

\subsection{Compaction time\label{sec-relax-time}}
\begin{figure}[t!]
\begin{center}
\includegraphics*[scale=0.32]{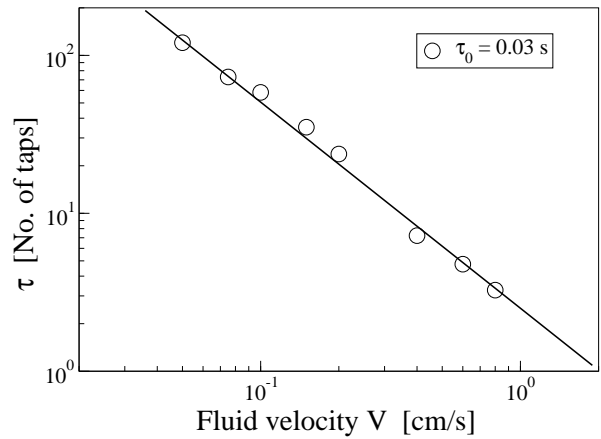}
\end{center}
\caption{\label{f_tau_V} The relaxation time $\tau$ increases with a power law (Eq.~\ref{eq-tau-V}) as
the fluid velocity decreases.
}
\end{figure}

The relaxation time $\tau$ which appears in Eq.~\ref{eq-se} is a measure of the number of flow pulses
required by the system to reach stationarity. The experiments of Ref.~\cite{Swinney05} have investigated a
range of parameters $V,\tau_0$ where the system reaches a stationary state 
after few flow pulses. On the contrary, the experiments of Bideau et. al.~\cite{Philippe02} have
investigated a range of parameters where the compaction dynamics of the system is glassy like. They report
a relaxation time following an Arrhenius behavior with the inverse maximum acceleration of the pack,
$\tau \propto \exp\left(\Gamma^{-1}\right)$. In the system under investigation here the relaxation time
also evidences the existence of a glassy dynamics of the system, as it diverges with a power law with
decreasing fluid velocity,
\begin{equation}
\label{eq-tau-V}
\tau \propto V^{-\beta}
\end{equation}
with $\beta \simeq 1.17 \pm 0.04$, as shown in Fig.~\ref{f_tau_V}
(for data with tap length $\tau_0 = 0.03$ s). The same behavior is
observed for different values of the tap length $\tau_0$.

\subsection{Stationary dynamics}
\begin{figure}[t!]
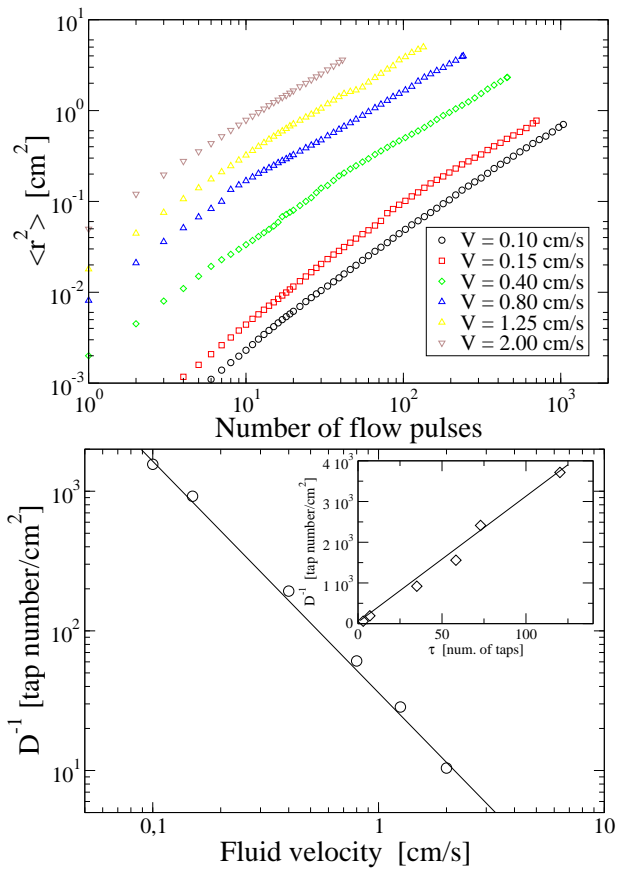

\begin{center}
\includegraphics*[scale=0.32]{msd_tap.eps}
\includegraphics*[scale=0.32]{diff_coeff.eps}
\end{center}
\caption{\label{fig_diffusion} (Color online) Left panel: mean square displacement (in the stationary
state) for $\tau_0=0.03$ s and fluid velocities (from top to bottom) $V=2.00,1.25,0.8,0.40,0.15$
and $0.10$ cm/s. Right panel: the inverse diffusion coefficient (measured for those points in which the
diffusion regime is attained) diverges as a power law, $D^{-1} = a V^{-\beta}$ as the fluid velocity $V$
decreases. Inset: Stokes-Einstein relation between the compaction time and the diffusion coefficient at
stationarity.}
\end{figure}
After having applied a long sequence of pulses up to reach the stationary state, we have computed the
mean square displacement,
\begin{equation}
\langle r^2(t) \rangle = \frac{1}{N}\sum_i^N ({\bf r}_i(t+t_w)-{\bf r}_i(t_w))^2,
\end{equation}
with ${\bf r}_i(t)$ position of grain $i$ after $t$ taps, and $t_w$ waiting time which depends on the
driving conditions ($t_w \simeq 3 \tau$). From the mean square displacement, shown in Fig.~
\ref{fig_diffusion}, we have extracted the diffusion coefficient $D$ ($\langle r^2(t) \rangle \propto Dt$)
which decreases with the fluid velocity as a power law, $D \propto V^\beta$, with the same exponent
observed for the dependence of the relaxation time on the fluid velocity (Fig.~\ref{fig_diffusion}).
This suggest the existence of 
relation $\tau \propto D^{-1}$ relating the compaction
time and the diffusion coefficient at stationarity, which is illustrated in the inset of
Fig.~\ref{fig_diffusion} (lower panel).


The mean square displacement evidences the absence of a subdiffusive regime in the slow dynamics of
granular media subject to flow pulses, a signature of the cage effect in supercooled liquids. In
Sec.~\ref{sec-st} we will show that particles may be constrained in a cage, but that the escaping time
is too small (few taps) in order to affect the mean square displacement.

\section{Structure evolution\label{sec-str}}
\subsection{Radial distribution function\label{sec-gr}}
The radial distribution function $g(r)$ is the probability distribution of finding the center of a
particle in a given position at a distance $r$ from a reference sphere. Since it contains information
about long range interparticle correlations, it is a common tool in the characterization of packing
structures.
Here we study how the radial distribution functions, which we normalize as usual is such a way that
$g(r) \to 1$ for $r \to \infty$, evolves during compaction.
\begin{figure}[t!]
\begin{center}
\includegraphics*[scale=0.32]{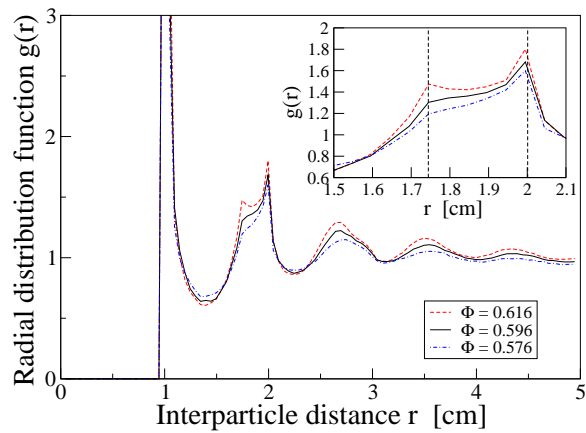}
\end{center}
\caption{\label{f_gr} (Color online) Radial distribution function for packing of volume fraction
$\Phi = 0.576, 0.596$ and $0.616$. As the volume fraction increases we observe a small increment of the
peaks at $\sqrt 3$d and $2$d, with $d=1$cm grain diameter.
}
\end{figure}

Fig.~\ref{f_gr} (main frame) shows $g(r)$ at different times during
compaction ($\tau_0 = 0.03$~s, $V = 0.2$~cm/s); similar results are
found with different values of $\tau_0$ and $V$. The first strong
peak at $r = d = 1$~cm corresponds to the high probability of a
having a neighbor in contact. This peak characterizes all dense
systems of hard particles, as a consequence of their
impenetrability. In a granular media at rest under gravity it is
enhanced by the fact that, in order for the system to be stable,
each grain must contact other grains. The following two maxima, as
shown in the inset, appears at $r = \sqrt{3}d$ and $r =2d$.
Both of them increase with the volume fraction indicating an
increasing organization of the packing (not necessarily related to the formation
of ordered structures~\cite{Aste}). A similar dependence
of the secondary peaks of the radial distribution function on the
volume fraction has been observed experimentally by T. Aste {\it at
al.}~\cite{Aste}, which have investigated via x-ray tomography packs
with different densities, and numerically by L.E. Silbert {\it at
al.}~\cite{Silbert1}.

\subsection{Vorono\"{\i} tessellation\label{sec-voronoi}}
In a granular pack of monodispere spheres of volume fraction $\Phi$ the mean volume occupied by a particle
is $V_p/\Phi$, where $V_p$ is the volume of a particle. When the system is in a disordered state there
will be both particles occupying a larger volume, and particles occupying a smaller one. It is
therefore instructive to investigate what is the probability that a given particle occupies a volume $v$,
and how this probability changes during compaction.
\begin{figure}[t!]
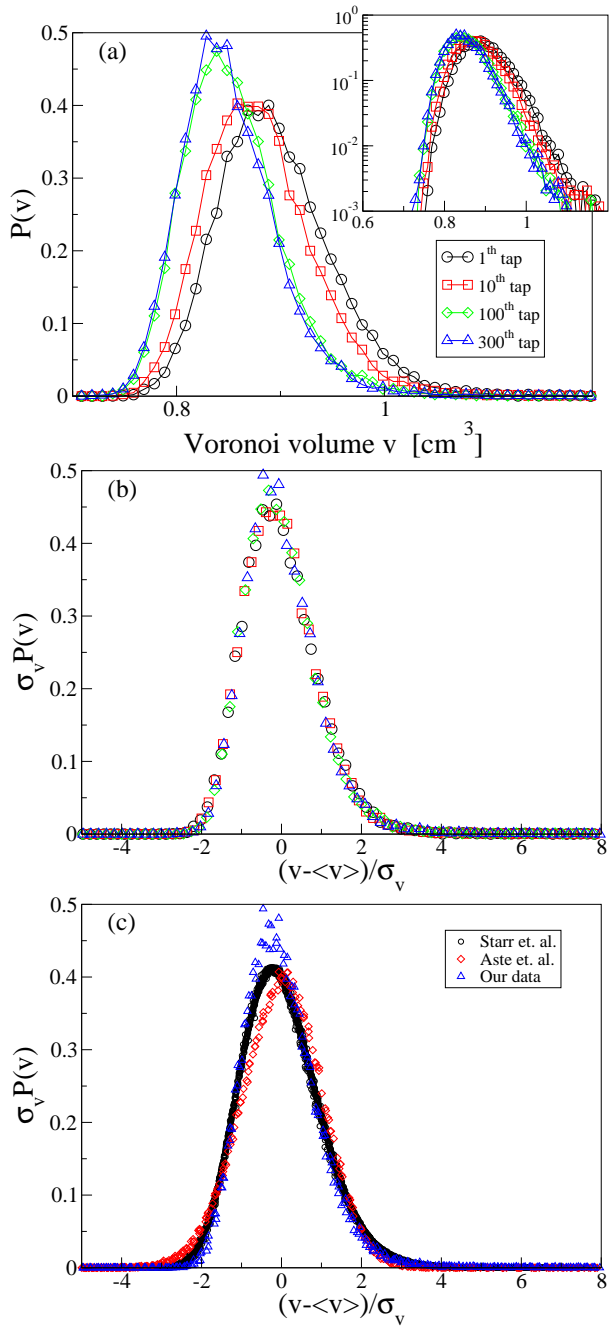

\begin{center}
\includegraphics*[scale=0.32]{voronoi_volumes.eps}
\includegraphics*[scale=0.32]{scaling_voronoi.eps}
\includegraphics*[scale=0.32]{compare_scaling.eps}
\end{center}
\caption{\label{f_voronoi} (Color online) (a) Distribution of the Vorono\"{\i} volumes in the sample after $1,10,100$ and $300$ taps. As the system compactifies the standard deviation $\sigma_v$ and the
mean value $\langle v \rangle$ of the distribution decrease. (b) Scaling of the distributions of the Vorono\"{\i} volumes shown in Panel (a). Same symbols are used. (c) The same scaling has been found in MD simulations of a model of glass-formes
by Starr {\it et al.}~\cite{Starr02}, and is verified by the experimental data of Aste {\it et al.}~
\cite{Aste}. However data from different sources do not scale on the same master curve.}
\end{figure}

To this end one has to operatively define what is the volume occupied by a particle: by using the
Vorono\"{\i} tessellation (as in~\cite{Aste}) we define the volume $v_i$ occupied by particle $i$
as the volume of the convex polyhedron which contains all points closer to particle $i$ than to any other
particle.
Fig.~\ref{f_voronoi}.a shows the distribution $P(v)$ of the Vorono\"{\i} volumes of a system tapped
with $\tau_0=0.03$~s and $V= 0.2$~cm/s, after $1,10,100$ and $300$ taps. These are slightly asymmetric
distributions with exponential tails (Fig.~\ref{f_voronoi}.a, inset). The asymmetry is a standard feature
of the Vorono\"{\i} distribution: for an ideal gas in one dimension, for instance,
$P(v) \propto v \rho^2 (1-\rho)^v$, where the Vorono\"{\i} volume $v$ of a given atom is half of the
distance between its left and right nearest neighbors.
As the system compactifies both the mean value $\langle v \rangle$ and the standard deviation $\sigma_v$
of the distribution decrease. However the distribution retains its functional form:  when $\sigma_v P(v)$
is plotted versus $(v-\langle v \rangle)/\sigma_v$ all of the different curves scale on the same master
curve, as shown in Fig.~\ref{f_voronoi}.b. This scaling suggests the existence of a single geometrical
structure of the system, only specified by its volume fraction, and tell us that there are no dramatic
structural changes during compaction.

The same scaling of the distribution of Vorono\"{\i} volumes has
been observed by F.W. Starr {\it et al.}~\cite{Starr02} in MD
simulations of a glass-forming polymer melt, and is verified by the
experimental data on granular packs of T. Aste {\it et
al.}~\cite{Aste}. However the data from these different sources do
not scale on the same muster curve, as shown in
Fig.~\ref{f_voronoi}.c. The discrepancy which may be due to the fact
that Ref.~\cite{Starr02} investigates a thermal system, while
Ref.~\cite{Aste} investigates a polydisperse system, needs further
investigation.


\subsection{Force distribution\label{sec-force}}
\begin{figure}[t!]
\begin{center}
\includegraphics*[scale=0.32]{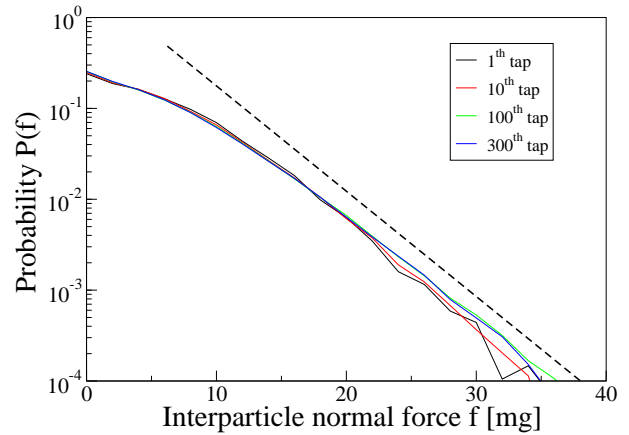}
\end{center}
\caption{\label{f_force} (Color online) The distribution of
interparticle normal forces decay exponentially at high forces, it
appears to be insensitive on the packing fraction of the system.}
\end{figure}
Since the work of Mueth {\it et. al.}~\cite{Mueth} interparticle force distributions have become a
standard tool for the characterization of granular packs. There are now experiments, numerical
simulations and theories (see~\cite{Snoeijer04} and references therein) finding an exponential
decay at high forces. This exponential decay is the signal of an heterogeneous structure of the system:
while most of the interparticle normal forces have magnitude close to the mean value, there are also
interparticle normal forces of much higher magnitude. Moreover, these high forces have been shown to
be spatially related, giving rise to the well know force chains.


We have studied the evolution of the probability distribution of normal forces during compaction. In
order to avoid the effect of gravity (due to the absence of vertical walls the mean vertical stress
depends on the depth) we have computed the probability distribution of the normal forces between grains
contacting in a point at a height $z$ enclosed in a thin horizontal slice ($z > 8$ cm and $z < 10$ cm).
Similar results are observed when selecting different horizontal slices of our system. Fig.~\ref{f_force}
shows the force probability distribution after $1,10,100$ and $300$ taps (corresponding to volume fractions in the range $0.576-0.616$). Since the distributions collapse on the same curve normal forces appears not to be affected by the density of the system (in the range we have investigated). We conclude that the force distribution is rather insensitive to the density of the granular pack, as also observed in Ref.~\cite{Blair01} and in~\cite{hern} (where it is show that forces do not couple with the density for thermal systems near the glass transition as well). 
It should be noted, however, that the use of the more realistic Herzian (instead of the linear) grain-grain interaction model (see Sec.~\ref{sec-model}) may change the properties of the force distribution.

\subsection{Force-Volume correlations}
\begin{figure}[t]
\begin{center}
\includegraphics*[scale=0.32]{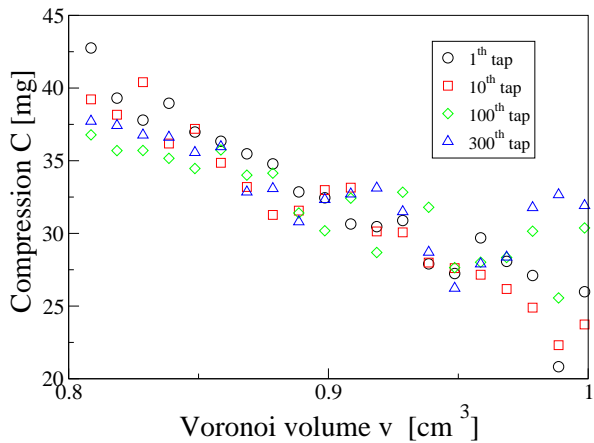}
\end{center}
\caption{\label{f_corr_vol_for} (Color online) The compressional force $C$ of a particle decreases
with its Vorno\"i volume.
}
\end{figure}
The relation between geometrical structure of the packing and interparticle forces has been investigated
in a number of previous works (see~\cite{Snoeijer04} and reference therein). Here we report on a relation between the Vorono\"{\i} volume associated to a particle and the forces acting on it. To this end we define
the compressional force $C_i$ acting on particle as:
\begin{equation}
\label{eq-compression}
C_i = \sum_{i\neq j} |\vec f_{ij}|
\end{equation}
where $|\vec f_{ij}|$ is the normal force of interaction between
particles $i$ and $j$. $C_i$ measures how much particle $i$ is
compressed as the pressure acting on it is $C_i/S_i$, where $S_i =
4\pi (d/2)^2$ is the particle surface. Fig.~\ref{f_corr_vol_for}
shows that the compressional force decreases with the
Vorono{\"i} volume.

A simple explanation of the decreasing of the compressional force with
the Vorono{\"i} volume can be obtained via the following argument.
Consider two contacting particles at a distance $l$. Their
interparticle force decreases with $l$ and particularly in our case,
due to the computational model used, $f = k(d-l)$, where $d$ is the
diameter of a particle. On the other hand the Vorono{\"i} volume $v$
of one of these particles increases with $l$. Assuming $v \propto
l^\alpha$, one obtains $f \propto (d-v^{1/\alpha})$, i.e. a decrease of the compressional 
force with the Vorono{\"i} volume. The data of Fig.~\ref{f_corr_vol_for}
are consistent with $\alpha = 3$, as expected for dimensional reasons.
However a reliable estimate of $\alpha$ is difficult to obtain as 
Vorono\"i volumes vary in a small range.


\section{Grain motion\label{sec-st}}
In this section we investigate how a granular pack moves during a
single tap. We consider a system subject to a tap dynamics with
$\tau_0 = 0.03$ s and $V = 0.2$ cm/s, and compare the state reached
after the application of $n$ taps, with the state reached after
the application of one more tap (with $n = 1,10,100,300$).

Particularly we investigate the distribution of particle
displacements, the heterogeneity of particle motion, and the
correlation between a particle displacement and both its
Vorono\"{\i} volume and the compressive force acting on it.
\begin{figure}[t]
\begin{center}
\includegraphics*[scale=0.8]{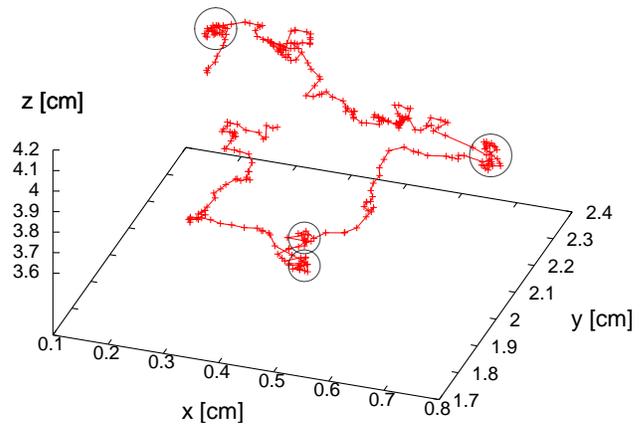}
\end{center}
\caption{\label{f_trai} (Color online) Typical trajectory of a
particle during compation, followed for $300$ taps during compaction
($V = 0.2$ cm/s, $\tau_0 = 0.03$ s). Circles illustrate that
sometimes a particle is confined in cages formed by its neighbors. }
\end{figure}

\subsection{Particle trajectory and cage motion}
Fig.~\ref{f_trai} shows a typical trajectory of a particle during
compaction. As observed in colloidal systems the trajectory is
characterized by period of time in which the particle is confined in
cages formed by its neighbors. Cage motion has also been observed before
in experiment of granular materials subject to continuous vibrations~\cite{Warr}
and to shear~\cite{pouliquen,requested_by_ref_1,requested_by_ref_2}.

The typical linear size of the cage
is roughly $0.05 d$, where $d = 1$ cm is the diameter of the
particles. A similar ratio between cage size and particle size has
been observed in~\cite{Weeks00}. However there is an important
qualitative difference between the motion of a particle in colloidal
suspension and other glass forming systems, and that observed in our
system. In glass forming systems a particle spend most of its time
rattling inside a cage, from which it escape via  infrequent
cage-breaking rearrangements. Here we observe the opposite
behavior: particles usually diffuse, and sometimes they get trapped
in a cage.
\begin{figure}[t!!]
\includegraphics*[scale=0.32]{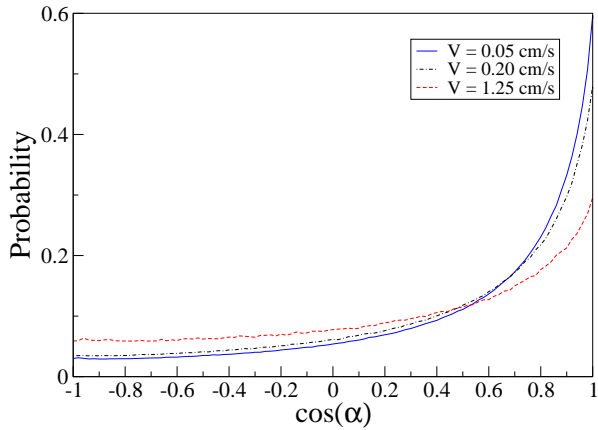}
\caption{\label{f_cosa} (Color online) Probability distribution of
$\cos(\alpha)$, where $\alpha$ is the angle formed by the
displacements $\vec \Delta$ of a particle in two subsequent taps.
The distribution evidences that particles usually travel along
straight lines, and that this tendency increases as the driving
intensity decreases. }
\end{figure}
This unusual property of the trajectory is put in evidence by the
study of the distribution of the angle $\alpha$ formed by the
displacement $\vec \Delta_n$ of a particle during tap $n$, and the
displacement $\vec \Delta_{n+1}$ of the same particle during the
following tap. Figure.~\ref{f_cosa} shows the distribution of
\begin{equation}
\cos(\alpha) = \frac{\vec \Delta_n \cdot \vec \Delta_{n+1}}{|\vec \Delta_{n+1}||\vec\Delta_{n+1}|},
\end{equation}
which is strongly peaked near $1$. This is a clear indication that
particles usually move along straight lines (as also confirmed by
Fig.~\ref{f_trai}), an that cage motion is negligible: for a
particle rattling in a cage two consecutive displacements have
opposite direction, and $\cos(\alpha) \simeq -1$.

The absence of cage motion is due to the driving mechanism. During a tap the
system expands, cages break, and it is easier for the grains to move one
with respect to the other.

\subsection{Particle displacements}
\begin{figure}[t!]
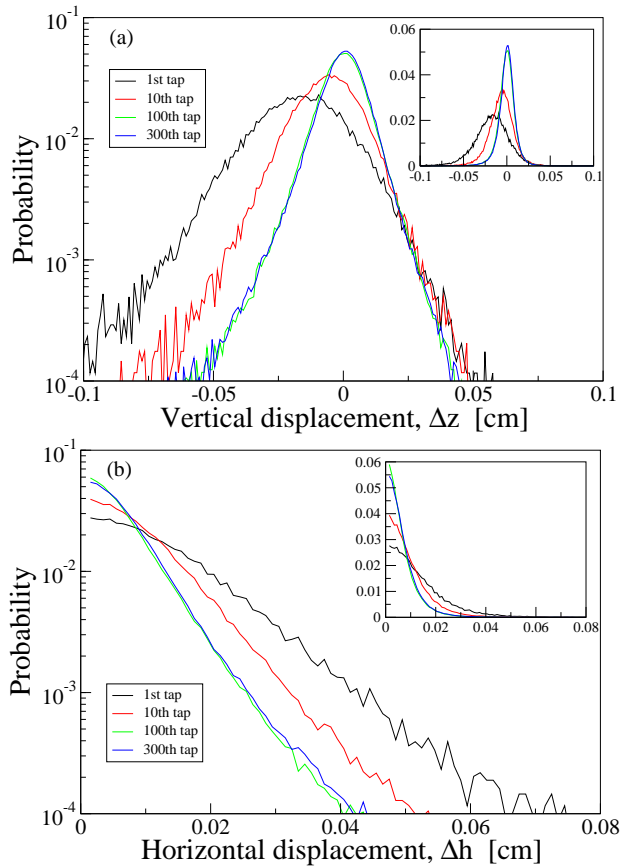

\begin{center}
\includegraphics*[scale=0.32]{p_dz.eps}
\includegraphics*[scale=0.32]{p_do.eps}
\end{center}
\caption{\label{f_pd} (Color online) Probability distribution of
vertical ($\Delta z$, upper panel) and horizontal ($\Delta h$, lower
panel) displacements of a grain during the $1$st, $10$th, $100$th
and $300$th tap in semi-logarithmic (main panels) and linear
(insets) scale. }
\end{figure}

Here we investigate the evolution of the distribution of particle
displacements during a single tap. After the application of a flow
pulse to our system (and the following relaxation) a particle $i$,
initially located in $\vec r_i$, will be in a new position position
$\vec r_i + \vec \Delta_i$, where $\vec \Delta_i$ denotes its
displacement. We examine below the probability that a particle makes
a given displacement $\vec \Delta$. Due to presence of gravity,
which breaks the up-down symmetry of the system, it is convenient to
separate the vertical component of the displacement, $\Delta z_i$,
from the horizontal ones, $\Delta x_i$ and $\Delta y_i$. As $\Delta
x_i$ and $\Delta y_i$ have the same, even distribution, we have
studied the evolution of the distribution of $\Delta h = |\Delta x|$
($|\Delta y|$). In order to follow the dynamics of the system, in
Fig.~\ref{f_pd} we have plotted the probability distribution of
$\Delta z$, $P_z(\Delta_z)$, (upper panel) and of $\Delta h$,
$P_h(\Delta_h)$, (lower panel) during tap number $1,10,100$ and
$300$.

During the compaction process displacements with $\Delta z < 0$ are
more probable than those with $\Delta z > 0$. Therefore, as shown in
Fig.~\ref{f_pd} (upper panel), $P_z(\Delta_z)$ is asymmetrical. The
asymmetry of the distribution decreases as the system compactifies
and after $300$ taps, when stationarity is almost attained,
$P_z(\Delta_z)$ appears to be nearly symmetric. Accordingly the
value of $\Delta z$ where $P_z(\Delta_z)$ has its maximum increases,
starting from a negative value, until it reaches zero. Also, it is
apparent that as the system compactifies the variance of the
distribution decreases. The probability of a large vertical
displacement to occur is smaller in a dense rather than in a fluffy
system. This is also true for the probability of large horizontal
displacements, as shown in Fig.~\ref{f_pd} (lower panel).

An important feature of the probability of both vertical and
horizontal displacements is the nearly exponential decay at large
displacements. This is an indication of the fact that, while during
a tap most of the particles are subject to small displacements, very
few of them may move much more. We conclude that the system is
characterized by an heterogeneous dynamics. This is a well know
property of thermal systems, like liquids or colloids
(see~\cite{Ediger00} for a review), which appears upon cooling the
system near the glass transition. For instance, in colloidal systems
the exponential tail of the particle displacement distribution has
been experimentally observed by Weeks {\it et. al.}~\cite{Weeks00}.

\begin{figure}[t]
\begin{center}
\includegraphics*[scale=0.35]{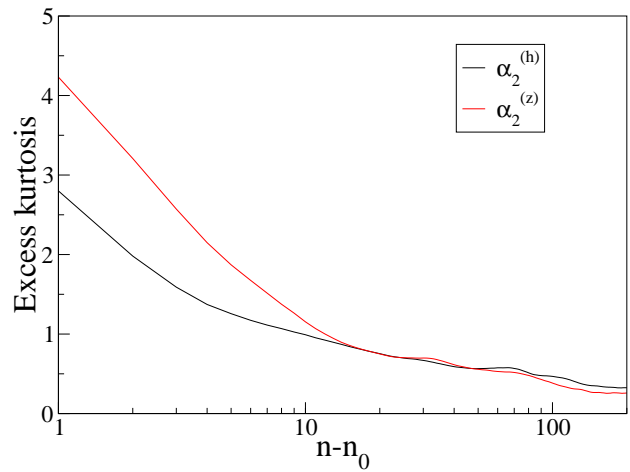}
\end{center}
\caption{\label{fig_ex_kur} Time variation of the excess kurtosis of
the probability distributions of the total horizontal ($\alpha^h_2$)
and vertical ($\alpha^z_2$) displacements of a grain since tap $n_0 = 100$.
The monotonic decay evidences the marginal role played by cage motion in our system.
}
\end{figure}

The probability distribution function of particles displacements can be
further analyzed via the study of the excess-kurtosis
\begin{equation}
\alpha^{(i)}_2 = \frac{\langle (\Delta_i-\langle\Delta_i\rangle)^2 \rangle^2}{3\langle(\Delta_i-\langle\Delta_i\rangle)^4\rangle}-1,~~,i=h,z,
\end{equation}
a comparison between the second and the fourth central moment of the distribution, which is zero for
a gaussain distribution. We have first considered the excess-kurtosis as obtained from
the probability of particle displacements in a single tap: $\alpha^h_2$ and $\alpha^z_2$
fluctuate from tap to tap, and their mean values are $\alpha^h_2 = 2.8 \pm 0.1$, $\alpha^z_2 = 4.3 \pm 0.2$.
This positive values indicate that the probability distribution of
particle displacements is more peaked with respect to a gaussian distribution.
Then, we have considered the excess-kurtosis of the probability distribution
of the horizontal and vertical components of $\vec r_i(n) - \vec r_i(n_0)$, the total displacement of a grain 
after tap $n_0$, where $n_0 = 100$ correspond to the compaction time. 
As expected form the central limit theorem at long times this excess-kurtosis is zero:
Fig.~\ref{fig_ex_kur} shows, interestingly, that in our system the excess kurtosis 
approach zero with a monotonic decay. This is in sharp contrast with the observations
made in glass forming system, as in Ref.~\cite{Weeks00}, where a peak is observed at the
$\alpha$ relaxation time, i.e., when cage rearrangements occur. Therfore Fig.~\ref{fig_ex_kur}
confirms the marginal role played by cage motion in our system.

Our results put in evidence a very smooth behavior of the particles
displacement distribution. Particularly we have not observed any
`rare event' (displacement of the order of $0.4$ particle
diameters), recently observed by Ribi{\`e}re and
coworkers~\cite{Ribiere} in the study of a granular system
undergoing compaction. This is probably due to the different driving mechanism
of the systems. In shaken system, in fact, grains move one with respect to the other
mainly when the pack settle downs and a shock wave propagates
upwards; in our system, on the contrary, there is not a shock wave propagating
as grains settle down slowly, and relative grain motion occurs during the tap.


\subsection{Spatial heterogeneous dynamics}
\begin{figure*}[ht!!]
\begin{center}
\includegraphics*[scale=1]{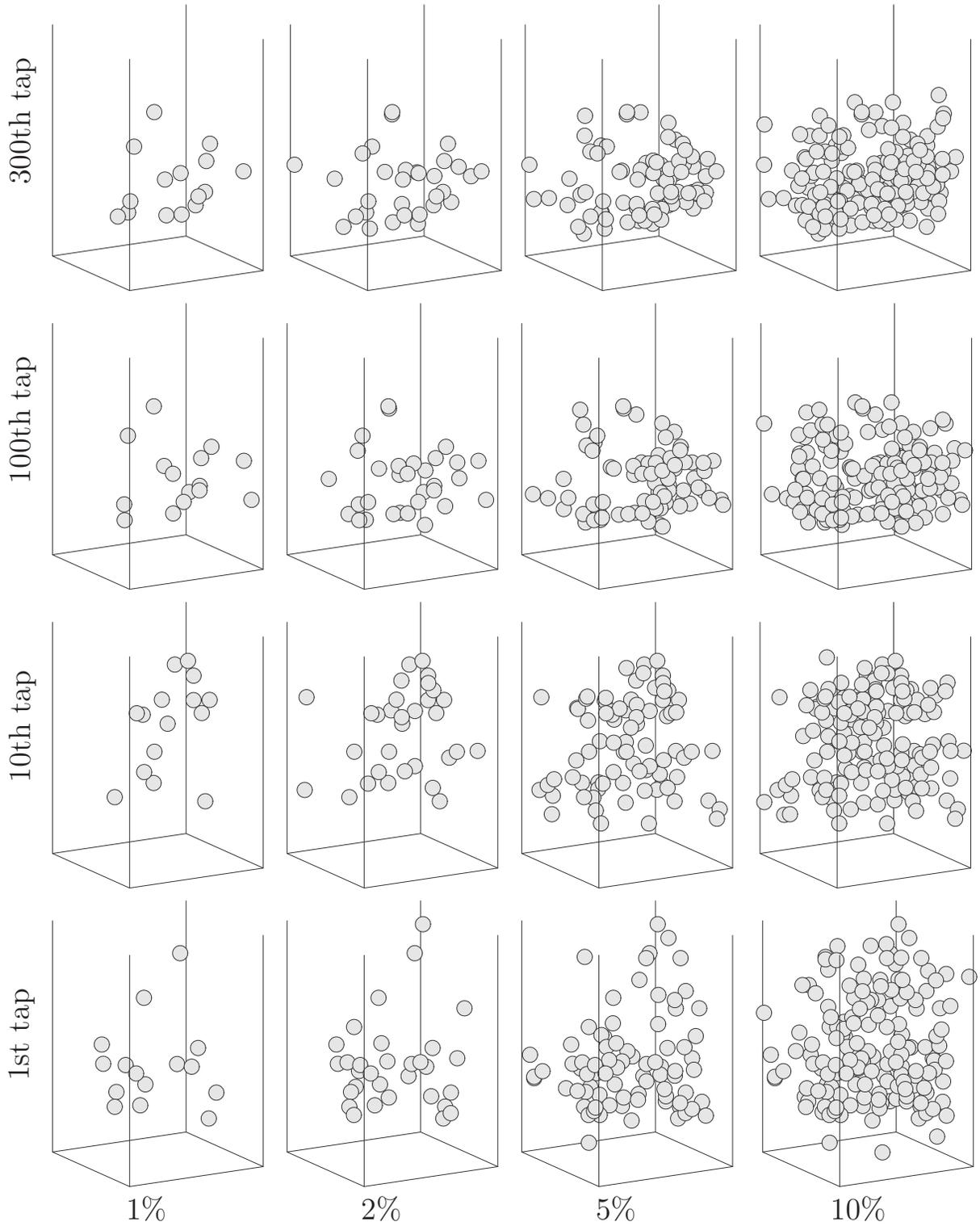}
\end{center}
\caption{\label{f_etero} Granular compaction is characterized by a
spatially heterogeneous dynamics. After the $k$th tap ($k =
1,10,100,300$, and volume fraction $\Phi = 0.576,0.596,0.612,0.616$) 
we plot the position of the $n_p$ particles which
have experienced the largest displacement during the last tap, with
$n_p = pN$ and $p = 1\%, 2\%, 5\%$ and $10\%$. }
\end{figure*}

In supercooled liquids and dense colloidal systems the dynamics is
heterogeneous as there are both slow and fast particles. Moreover,
fast particles are know to be spatially correlated as they appear to
form clusters~\cite{Ediger00,Weeks00,Glotzer00}. Here we show that
the exact same tendency characterizes a granular system subject to
vertical taps, as suggested in Ref.s~\cite{Arenzon,Lefevre}.

In order to characterize the spatially heterogeneous
dynamics~\cite{Kob97,Glotzer00,Glotzer04} one usually resorts to a
four-point time-dependent density correlation function and to its
fluctuations (susceptibility). This latter  measures the correlated
motion between pairs of particles. As this motion is decorrelated on
short and long times, the susceptibility show a well defined maximum
at a given time. Unfortunately we cannot follow this line here as
many averages are needed in order to get clear data on the
fluctuations, and our simulations are computationally too expensive.
Moreover during compaction the system is not in a stationary state.
In order to measure the degree of spatally heterogeneous dynamics we
have therefore used a different method, based on the comparison
between our system and a random one, as discussed below.

We apply $k$ flow pulses ($k=1,10,100,300$) to our system (who reaches
volume fraction $0.576,0.596,0.612,0.616$) and we determine the $n_p$
particle having experienced the largest displacement during the last tap.
Here $n_p = p N$, were $N= 1600$ is the total number of particles, 
$p = 1\%, 2\%, 5\%$ and $10\%$. 
When these fast particles are drawn, as in Fig.~\ref{f_etero},
it is apparent that they form clusters, an clear evidence of the 
spatially heterogeneous dynamics. We have quantified the degree of
heterogeneity of the system as follows. After every tap we have
determined the $n_p$ faster particles of the system, and determined the number 
$s_p$ of couples of these particles made of neighboring particles (we consider two
particles to be neighbor if the distance between their center is
smaller than $1.2$ particle diameters). Then we have computed the
same quantity $s^{\rm random}_p$ in the case of $n_p$ randomly
selected particles of the system. A measure of the degree of spatial
heterogeneous dynamics is given by
\begin{equation}
\label{eq-hetero}
\xi_p = \frac{s_p-s^{\rm random}_p}{s^{\rm random}_p}.
\end{equation}

\begin{figure}[t]
\begin{center}
\includegraphics*[scale=0.32]{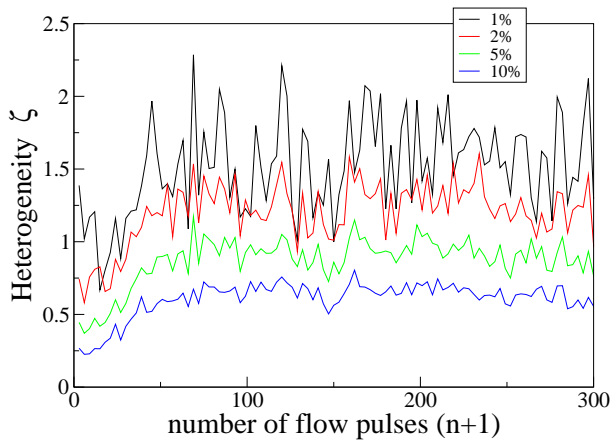}
\end{center}
\caption{\label{f_etero_2} (Color online) Granular compaction is
characterized by a spatially heterogeneous dynamics. This is
quantified by the parameter $\xi_p$ (see Eq.\ref{eq-hetero}) which
is plotted here as a function of the number of taps for $p = 1\%,
2\%, 5\%$ and $10\%$ (from top to bottom).}
\end{figure}

Clearly $\xi_p \simeq 0$ for an homogeneous systems, $\xi_p > 0$ if
faster particles form clusters, while $\xi_p <0$ if faster particles
tend to be apart. Fig.~\ref{f_etero_2} shows the evolution of
$\xi_p$ with the number of taps. In all cases $\xi_p > 0$, signaling
the presence of an heterogeneous dynamics. When the system 
approaches the steady state, and compaction stops, 
$\xi_p$ fluctuates around a plateau which varies with $p$ between
$0.5$ and $1.5$. These are very high values, indicating a high
degree of spatial heterogeneity of the system.


\subsection{Volume-Displacement correlation}
\begin{figure}[t]
\begin{center}
\includegraphics*[scale=0.32]{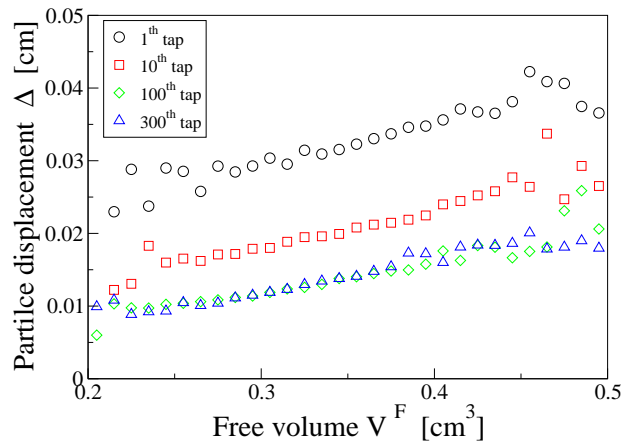}
\end{center}
\caption{\label{f_corr_vol_dis} (Color online) The mean displacement
$\Delta$ of particle during a single tap increases with its Vorono\"{\i} volume $v$. The displacements are measured during tap $1,10,100$ and $300$,
when the volume fraction of the system is $0.576,0.596,0.612,0.616$.}
\end{figure}

It is well known than many equilibrium and transport properties of
dense fluids depend on the space available for molecular motion. For
instance the well known {\it free volume} theory developed by Choen
and Turnbull~\cite{Choen59} to explain the divergence (with a
Vogel-Tamman-Fulcher law) of the relaxation time of many glass
formers as the temperature is decreased, is based on the idea that
the space available for molecular motion decreases with the
temperature. It is therefore interesting to check for correlations
between the displacement of a particle and its free volume in our
granular system. There are many possible way to define the free
volume of a particle. Here we approximate, for simplicity sake, we consider
the free volume of particle $i$ to be $V_i^F=v_i-V_0$,
where $v_i$ is the Vorono\"{\i} volume of the particle (see Sec.~\ref{sec-voronoi})
before the application of a tap, and $V_0 = 4/3\pi (D/2)^3$ its volume.
The displacement $\Delta_i$ of particle $i$ ($\Delta_i = (\Delta_x^2
+ \Delta_y^2 + \Delta_z^2)^{1/2}$) is the distance between the
position of the particle before and after the application of the
tap.

A possible connection between $V_i^F$ and $\Delta_i$ is suggested by
the similarity between the the probability distribution functions of 
Vorono\"i volumes (see Fig.~\ref{f_voronoi}.a), and that of particles displacemets 
(Fig.~\ref{f_pd}). Both of them have an
exponential tail at high values. Moreover they evolve in a
qualitatively similar way (the variance and the mean value decrease)
as the system compactifies.

In order to test this possible correlation we have computed during
the $n$th tap ($n = 1,10,100,300$) ($\tau_0 = 0.03$ s, $V = 0.2$
cm/s) the mean value of the displacement $\Delta$ of all the
particles with free volume $V^F$. The dependence of $\Delta$
on $V^F$ is shown in Fig.~\ref{f_corr_vol_dis}. This figure puts in
evidence the existence of an  almost linear correlation between
Vorono\"{\i} volumes and particle displacements, with $\partial
\Delta/\partial v \simeq 0.01/0.2 = 0.05$ cm$^{-2}$: the larger the
Vorono\"{\i} volume of a particle, the bigger its displacement.

\subsection{Force-Displacement correlation}
\begin{figure}[t!!]
\begin{center}
\includegraphics*[scale=0.32]{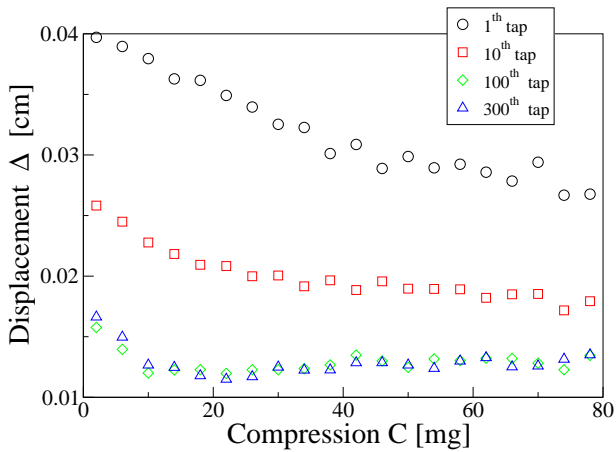}
\end{center}
\caption{\label{f_corr_for_dis} (Color online) The mean displacement
$\Delta$ of a grain as a function of its compressional force. When
the packing is loose grains with higher compressional forces move
less. As the granular pack compactifies the displacement of a
particle appears not to depend on its compression. }
\end{figure}

In the previous paragraph we have shown that there is a correlation
between the displacement $\Delta$ of a particle during a single tap,
and its free volume. Fig.~\ref{f_corr_for_dis}
investigates the correlation between displacement and compression
$C$ (see Eq.~\ref{eq-compression}) of a particle. The figure puts in
evidence that for small values of the compressional force there is a
decreasing linear relation between displacement and compression of a
particle, while for higher value of the compressional force the two
becomes uncorrelated.


\section{Conclusions\label{sec-conc}}
We have reported results of a numerical simulations of a packing of
monosize spheres submitted to vertical taps made of flow pulses as
in the experiment of Ref.~\cite{Swinney05}. Our results relative to
the dynamics of the systems confirms earlier experimental
observation: as the intensity of vibration decreases both the value
of the volume fraction reached at stationary and the compaction time
increases~\cite{Swinney05,Nowak97,Nowak98,Ben-Naim,Philippe02}. The
increase of the compaction time with the decreasing of the vibration
intensity is dramatic as this appears to diverge with a power law
when the fluid velocity goes to zero.

The analysis of the evolution of several structural quantities
during compaction has revealed that this is not accompanied by any
particular geometrical modification. The radial distribution
function and the Vorono\"{\i} volume distribution, for instance,
smoothly change as the density of the granular system increases.
In particular the collapse of the Vorono\"{\i} volume distributions
(sec.~\ref{sec-voronoi}) evidences the presence of a single
underlying geometrical structure in the system. Also, the
probability distribution function of normal forces between grains do
not change during compaction.

The analysis of the dynamics of compaction has revealed that this is
characterized by dynamical heterogeneities. The probability that,
during a tap, a particle makes a given displacement decreases
exponentially with its size, resembling observations made in dense
colloidal systems~\cite{Weeks00}. Similarly we have observed that,
during a tap, faster particle tend to form cluster, as observed both
in colloidal~\cite{Weeks00} and in
glass-forming~\cite{Ediger00,Glotzer00,Glotzer04} systems. There is,
however, a marked difference between a typical trajectory of a
granular particle during compaction, and a typical trajectory of a
particle in supercooled liquids. Precisely, in supercooled liquids a
particle spend most of time in cages formed by its neighbors, and
occasionally makes large displacement escaping from the cage. In our
system, instead, particles usually diffuse, and sometimes are
trapped in a cage. This different behaviour is due to the particular driving of our system
as in our system, when the the flow in on the system expands and the grains are able to escape
from their cages.

As the slowdown of the dynamics is related to the space available
for particle motion, we have studied the correlation between the
displacement of particle during a tap and its Vorono\"{\i} volume,
which is a rough estimate of its free volume. This analysis has
shown that particles with larger Vorono\"{\i} volumes are those who
make larger displacements during a tap.

\section*{Acknowledgements}
We thank A.~de~Candia, A.~Fierro, M.~Scr$\ddot{o}$ter and M.~
Tariza for helpful discussions,
F.W.~Starr and T.~Aste for sharing their data. Work supported by EU Network Number
MRTN-CT-2003-504712, MIUR-PRIN 2004, MIUR-FIRB 2001, CRdC-AMRA, INFM-PCI.


\begin{thebibliography}{50}
\bibitem{modello}
We use the model `L3' described in detail in Ref.~\cite{Silbert1}.
\end{thebibliography}
\end{document}